\documentclass{PoS}

\usepackage{cite}

\title{The sunrise integral and elliptic polylogarithms}

\ShortTitle{The sunrise integral and elliptic polylogarithms}

\author{Luise Adams\\
       PRISMA Cluster of Excellence, Institut f\"ur Physik,\\
      Johannes Gutenberg-Universit\"at Mainz,\\
      D - 55099 Mainz, Germany\\
        E-mail: \email{ladams01@uni-mainz.de}}

\author{\speaker{Christian Bogner}\\%
        Institut f\"ur Physik, Humboldt-Universit\"at zu Berlin,\\
	D - 10099 Berlin, Germany\\
       E-mail: \email{bogner@math.hu-berlin.de}}

\author{Stefan Weinzierl\\
       PRISMA Cluster of Excellence, Institut f\"ur Physik,\\
      Johannes Gutenberg-Universit\"at Mainz,\\
      D - 55099 Mainz, Germany\\
        E-mail: \email{weinzierl@uni-mainz.de}}

\abstract{We summarize recent computations with a class of elliptic generalizations
of polylogarithms, arising from the massive sunrise integral. For
the case of arbitrary masses we obtain results in two and four space-time
dimensions. The iterated integral structure of our functions allows
us to furthermore compute the equal mass case to
arbitrary order.}

\FullConference{Loops and Legs in Quantum Field Theory\\
                 24-29 April 2016\\
                 Leipzig, Germany}

\begin{document}
\section{Introduction}

Various methods for the symbolical computation of multi-loop Feynman
integrals rely on properties of classical polylogarithms 

\[
\textrm{Li}_{n}(z)=\sum_{j=1}^{\infty}\frac{z^{j}}{j^{n}},\,\,\left|z\right|<1,
\]
and their generalizations. By now the class
of multiple polylogarithms \cite{Gon2,Gon1}
\[
\textrm{Li}_{n_{1},...n_{k}}\left(z_{1},...,z_{k}\right)=\sum_{0<j_{1}<...<j_{k}}\frac{z_{1}^{j_{1}}...z_{k}^{j_{k}}}{j_{1}^{n_{1}}...j_{k}^{n_{k}}},\,\,|z_{i}|<1,
\]
is well-established in particle physics.

One of the advantages of these functions is their iterated integral
structure. For example a classical polylogarithm of weight $n\geq2$
can be written as 
\begin{eqnarray}
\textrm{Li}_{n}(z) & = & \int_{0}^{z}\frac{dx}{x}\textrm{Li}_{n-1}(x)\label{eq:Li_n integral}
\end{eqnarray}
and similar relations hold for the generalizations. 

One of the computational approaches making use of this property is
the method of differential equations \cite{Kot,Rem}. Here a Feynman integral 
is computed by integrating over a linear combination
of other Feynman integrals. If the latter are known
in terms of generalized polylogarithms and if they appear with integral kernels
in an appropriate set of differential forms, then the integral over
these expressions can be computed by use of relations such as eq.
\ref{eq:Li_n integral} and the result belongs to the same class of
functions. 

The computations summarized in these notes are motivated by the fact,
that multiple polylogarithms are not sufficient to express every Feynman
integral. We consider several cases of the massive sunrise integral,
which is a famous showcase of this problem. Various classes of functions
different from polylogarithms were applied to this integral in the
past. More recently, the case of equal masses in two space-time dimensions
was expressed with the help of an elliptic dilogarithm in \cite{BloVan}. 

We define a related class of elliptic generalizations of polylogarithms,
including a generalization depending on several variables. With the
help of these functions, we compute the sunrise integral in the case
of arbitrary masses at two and, with the help of dimension shift relations, near four space-time dimensions. We
furthermore show for the case of equal masses and two dimensions,
that all orders of the Laurent expansion can be expressed with the
help of our framework of functions. We provide an explicit algorithm
for the computation of these orders, relying on corresponding differential
equations and on the iterated integral structure of our class of functions.

\section{A class of elliptic generalizations}

We define a class of functions of variables $q,\, x_{1},...,x_{l},\, y_{1},...,y_{l}$.
They are related with polylogarithms and known elliptic generalizations.

\subsection{Definitions\label{sub:Definitions}}

For $l=1$ we define 
\begin{equation}
\textrm{ELi}_{n;m}(x;y;q)=\sum_{j=1}^{\infty}\sum_{k=1}^{\infty}\frac{x^{j}}{j^{n}}\frac{y^{k}}{k^{m}}q^{jk}\label{eq:ELi x case}
\end{equation}
and for $l>1$ we define 
\[
\textrm{ELi}_{n_{1},...,n_{l};m_{1},...,m_{l};2o_{1},...,2o_{l-1}}\left(x_{1},...,x_{l};y_{1},...,y_{l};q\right)
\]
\begin{eqnarray}
 & = & \sum_{j_{1}=1}^{\infty}...\sum_{j_{l}=1}^{\infty}\sum_{k_{1}=1}^{\infty}...\sum_{k_{l}=1}^{\infty}\frac{x_{1}^{j_{1}}}{j_{1}^{n_{1}}}...\frac{x_{l}^{j_{l}}}{j_{l}^{n_{l}}}\frac{y_{1}^{k_{1}}}{k_{1}^{m_{1}}}..\frac{y_{l}^{k_{l}}}{k_{l}^{m_{l}}}\frac{q^{j_{1}k_{1}+...+j_{l}k_{l}}}{\prod_{i=1}^{l-1}\left(j_{i}k_{i}+...+j_{l}k_{l}\right)^{o_{i}}}\label{eq:Funktionenklasse}
\end{eqnarray}
We will refer to these as ELi-functions.

By construction, this class of functions is closed under multiplication
with the $(l=1)$-case $\textrm{ELi}_{n;m}$ and under integration
over $\frac{dq}{q}.$ We have 
\[
\textrm{ELi}_{n_{1},...,n_{l};m_{1},...,m_{l};2o_{1},...,2o_{l-1}}\left(x_{1},...,x_{l};y_{1},...,y_{l};q\right)
\]
\[
=I^{o_{1}}\textrm{ELi}_{n_{1};m_{1}}(x_{1};y_{1};q')\textrm{ELi}_{n_{2},...,n_{l};m_{2},...,m_{l};2o_{2},...,2o_{l-1}}\left(x_{2},...,x_{l};y_{2},...,y_{l};q'\right)
\]

where $I^{o_{i}}$ denotes the $o_{i}$-fold integration 
\[
I^{o_{i}}=\int_{0}^{q}\frac{dq_{1}}{q_{1}}\int_{0}^{q_{1}}\frac{dq_{2}}{q_{2}}...\int_{0}^{q_{o_{i}-2}}\frac{dq_{o_{i}-1}}{q_{o_{i}-1}}\int_{0}^{q_{o_{i}-1}}\frac{dq'}{q'}\textrm{ for }o_{i}>0
\]
and $I^{0}=1.$ 

Combining the above ELi-functions, we furthermore define a class which
we will refer to as E-functions by
\[
\textrm{E}_{n;m}(x;y;q)=d_{n,m}\left(\frac{1}{2}\textrm{Li}_{n}(x)+c_{n,m}\frac{1}{2}\textrm{Li}_{n}(x^{-1})+\textrm{ELi}_{n;m}(x;y;q)+c_{n,m}\textrm{ELi}_{n;m}(x^{-1};y^{-1};q)\right)
\]
where $c_{n,m}=-1,$ $d_{n,m}=-i$ for even
$n+m$ and $c_{n,m}=1,$ $d_{n,m}=1$ for odd $n+m.$ 
We furthermore define
\[
\textrm{E}_{n_{1},...,n_{l};m_{1},...,m_{l};2o_{1},...,2o_{l-1}}\left(x_{1},...,x_{l};y_{1},...,y_{l};q\right)
\]
\[
=I^{o_{1}}\left(\textrm{E}_{n_{1};m_{1}}(x_{1};y_{1};q')-\textrm{E}_{n_{1};m_{1}}(x_{1};y_{1};0)\right)\textrm{ELi}_{n_{2},...,n_{l};m_{2},...,m_{l};2o_{2},...,2o_{l-1}}\left(x_{2},...,x_{l};y_{2},...,y_{l};q'\right).
\]
Our results for the sunrise integral, discussed below, will be expressed
in terms of E-functions and multiple polylogarithms.

\subsection{Relations with known functions}

In the case of all $o-$indices being zero, the ELi-functions are
products of the $(l=1)$-case:
\[
\textrm{ELi}_{n_{1},...,n_{l};m_{1},...,m_{l};0,...,0}\left(x_{1},...,x_{l};y_{1},...,y_{l};q\right)=\prod_{i=1}^{l}\textrm{ELi}_{n_i;m_i}(x_i;y_i;q).
\]
For $q=1$ the latter is furthermore just a product of polylogarithms due to 
\[
\textrm{ELi}_{n;m}(x;y;q)=\sum_{k=1}^{\infty}\frac{y^{k}}{k^{m}}\textrm{Li}_{n}(q^{k}x)\textrm{ and }\textrm{ELi}_{n;m}(x;y;1)=\textrm{Li}_{n}(x)\textrm{Li}_{m}(y).
\]

More notably, the E-functions are related to known versions of elliptic
polylogarithms. Let us briefly recall a basic principle behind such
functions. We consider a lattice of points $L=\mathbb{Z}+\tau\mathbb{Z}$
where $\tau\in\mathbb{C}$ with $\textrm{Im}(\tau)>0.$
A function of $x\in\mathbb{C}$ is called elliptic with respect to
$L$ if it is periodic under $x\rightarrow x+\lambda$ with $\lambda\in L.$
For a function $F$ of $z=e^{2\pi ix}\in\mathbb{C}^{\star}$ this
condition translates to
\[
F(z)=F(zq)\textrm{ for }q=e^{2\pi i\lambda},\,\lambda\in L.
\]
This concept
was first applied to define an elliptic dilogarithm in \cite{Blo}. Generalizations were introduced in \cite{Zag,BeiLev,Lev,GanZag}.

In \cite{BroLev} elliptic polylogarithms are defined as coefficients of the regular part of the Laurent expansion around $\alpha=0$ of functions
\begin{equation}
 E_m(z;u;q) = \sum\limits_{n \in {\mathbb Z}} u^n \textrm{Li}_m(q^n z)
\end{equation}
with $u=e^{2\pi i \alpha}$. The latter functions are related to the above functions $\textrm{E}_{n;m}(x;y;q).$ We
have for example 
\begin{eqnarray}
\textrm{E}_{2;0}(x;y;q) & = & \frac{1}{i}\left(E_{2}(x;y;q)-\frac{1}{2}\frac{1+y}{1-y}\zeta(2)-\frac{1}{4}\frac{1+y}{1-y}\ln^{2}(-x)\right.\nonumber \\
 &  & \left.-\frac{y}{(1-y)^{2}}\ln(-x)\ln(q)-\frac{1}{2}\frac{y\left(1+y\right)}{(1-y)^{3}}\ln^{2}(q)\right).\label{eq:relation between elliptic dilogs}
\end{eqnarray}
The functions $\textrm{E}_{n;m}(x;y;q)$ can furthermore be understood
as generalizations of Clausen- and Glaisher-functions, which are defined
by
\[
\textrm{Cl}_{n}\left(\varphi\right)=\frac{1}{2i}\left(\textrm{Li}_{n}\left(e^{i\varphi}\right)-\textrm{Li}_{n}\left(e^{-i\varphi}\right)\right),\,\textrm{Gl}_{n}\left(\varphi\right)=\frac{1}{2}\left(\textrm{Li}_{n}\left(e^{i\varphi}\right)+\textrm{Li}_{n}\left(e^{-i\varphi}\right)\right)
\]
for even $n$ and by
\[
\textrm{Cl}_{n}\left(\varphi\right)=\frac{1}{2}\left(\textrm{Li}_{n}\left(e^{i\varphi}\right)+\textrm{Li}_{n}\left(e^{-i\varphi}\right)\right),\,\textrm{Gl}_{n}\left(\varphi\right)=\frac{1}{2i}\left(\textrm{Li}_{n}\left(e^{i\varphi}\right)-\textrm{Li}_{n}\left(e^{-i\varphi}\right)\right)
\]
for odd $n.$ We have
\[
\textrm{lim}_{q\rightarrow0}\textrm{E}_{n;m}\left(e^{i\varphi};y;q\right)=\textrm{Cl}_{n}\left(\varphi\right)
\]
for $m$ being zero or even and
\[
\textrm{lim}_{q\rightarrow0}\textrm{E}_{n;m}\left(e^{i\varphi};y;q\right)=\textrm{Gl}_{n}\left(\varphi\right)
\]
for $m$ being odd.

\section{Cases of the massive sunrise integral\label{sec:The-massive-sunrise}}

The massive sunrise integral 
\[
S(D,\, t)=\int\frac{d^{D}k_{1}d^{D}k_{2}}{\left(i\pi^{D/2}\right)^{2}}\frac{1}{\left(-k_{1}^{2}+m_{1}^{2}\right)\left(-k_{2}^{2}+m_{2}^{2}\right)\left(-\left(p-k_{1}-k_{2}\right)^{2}+m_{3}^{2}\right)},
\]
which in various versions was considered by many authors \cite{Baileyetal,Bauetal,Bauetal2,Bauetal3,Beretal,BroFleTar,CafCzyRem,Caffoetal,CafGunRem,GroKoePiv,GroKoePiv2,LapRem,PozRem,RemTan,UssDav,RemTan2,Broadhu,KalKni,DavDel,DavSmi},
is a showcase for the mentioned problem, that there are Feynman integrals
which can not be expressed entirely in terms of multiple polylogarithms.
For arbitrary masses and arbitrary dimension $D,$ the integral was
computed in \cite{Beretal} in terms of Lauricella functions of type
C. The fact that none of the existing techniques provides a way to
expand these functions in terms of multiple polylogarithms so far
may be seen as a confirmation of the mentioned problem.

With respect to the variable $t=p^{2}$ which we consider in the region
$t\leq0,$ the integral $S(D,\, t)$ satisfies a differential equation
\[
L_{4}S(D,t)=T(D,t).
\]
Here $L_{4}$ is a differential operator of fourth order and the
inhomogeneous part $T(D,t)$ is a combination of tadpole integrals,
all of whose coefficients are polynomials
in $m_{1}^{2},\, m_{2}^{2},\, m_{3}^{2},\, t,\, D.$ In the following
we will consider coefficients in the Laurent series of $S(D,t),$
satisfying differential equations of fourth or lower order. These
coefficients will arise from the expansion at $D=2$ and at $D=4$
dimensions: 
\begin{eqnarray}
S(2-2\epsilon,t) & = & S^{(0)}(2,t)+S^{(1)}(2,t)\epsilon+\mathcal{O}\left(\epsilon^{2}\right),\label{eq:S(2)}\\
S(4-2\epsilon,t) & = & S^{(-2)}(4,t)\epsilon^{-2}+S^{(-1)}(4,t)\epsilon^{-1}+S^{(0)}(4,t)+\mathcal{O}(\epsilon).\label{eq:S(4)}
\end{eqnarray}

\subsection{The case of $D=2$ dimensions}

The case of exactly $D=2$ dimensions is a good starting point for
several reasons. Firstly, the Feynman integral is finite here. The
Laurent expansion in eq. \ref{eq:S(2)} begins with $S^{(0)}(2,t)$
which satisfies a differential equation \cite{MueWeiZay1}

\begin{eqnarray}
L_{2}\, S^{(0)}(2,t) & = & P(t),\label{eq:DiffEq y}
\end{eqnarray}
where $L_{2}$ is a second order differential operator with respect
to $t$ whose coefficients are polynomials in the squared masses and
$t.$ The inhomogeneous part $P(t)$ furthermore involves logarithms
of the squared masses.

Secondly, if we write the Feynman integral in terms of Feynman parameters,
the first Symanzik polynomial drops out in $D=2$ dimensions and the
integrand only involves the second one, which reads 
\[
\mathcal{F}=-x_{1}x_{2}x_{3}t+\left(x_{1}m_{1}^{2}+x_{2}m_{2}^{2}+x_{3}m_{3}^{2}\right)\left(x_{1}x_{2}+x_{2}x_{3}+x_{1}x_{3}\right).
\]
Even though we do not attempt to integrate out the Feynman parameters,
this polynomial plays an important role in our computations. The zero
set of this polynomial intersects the domain of the Feynman parametric
integral at three points in its corners. By choosing one of these
points as the origin, we obtain an elliptic curve defined by $\mathcal{F}.$
The corresponding Weierstrass normal form defines three zeros $e_{1},\, e_{2},\, e_{3}$
of the cubical equation. Using these as integration boundaries, one
canonically defines two period integrals $\psi_{1},\,\psi_{2}$ of
the elliptic curve. These evaluate to 
\[
\psi_{1}=\frac{4}{\tilde{D}^{\frac{1}{4}}}K(k),\;\,\psi_{2}=\frac{4i}{\tilde{D}^{\frac{1}{4}}}K(k')
\]
where 
\[
K(x)=\int_{0}^{1}dt\frac{1}{\sqrt{(1-t^{2})(1-x^{2}t^{2})}}
\]
is the complete elliptic integral of first kind and where 
\[
k=\sqrt{\frac{e_{3}-e_{2}}{e_{1}-e_{2}}},\, k'=\sqrt{1-k^{2}}=\sqrt{\frac{e_{1}-e_{3}}{e_{1}-e_{3}}}
\]
and 
\[
\tilde{D}=(t-(m_{1}+m_{2}-m_{3})^{2})(t-(m_{1}-m_{2}+m_{3})^{2})(t-(-m_{1}+m_{2}+m_{3})^{2})(t-(m_{1}+m_{2}+m_{3})^{2}).
\]
As $\psi_{1}$ and $\psi_{2}$ are solutions of the homogeneous equation
$L_{2}\, S^{(0)}(2,t)=0,$ the special solution of the inhomogeneous
eq. \ref{eq:DiffEq y} can be constructed by classical variation
of constants as an integral over a certain combination of the homogeneous
solutions. In this way, we obtain the full solution involving an integral
over complete elliptic integrals \cite{AdaBogWei1}. 

However, we find \cite{AdaBogWei2} that the solution can be written
alternatively as 
\begin{eqnarray}
S^{(0)}(2,t) & = & \frac{\psi_{1}(q)}{\pi}E^{(0)},\label{eq:result ellipt dilog}\\
E^{(0)} & = & \sum_{i=1}^{3}\textrm{E}_{2;0}(w_{i}(q);\,-1;\,-q)\label{eq:def E^(0)}
\end{eqnarray}
where $\textrm{E}_{2;0}$ is one of the E-functions. The dependence on $t$ is now given in
terms of $q$ which we define as $q=e^{\pi i\frac{\psi_{2}(t)}{\psi_{1}(t)}}$
in terms of the period integrals of our elliptic curve. The three
arguments $w_{1},$ $w_{2},$ $w_{3}$ are obtained explicitly from
the mentioned intersection points by transformations on the elliptic
curve. 

\subsection{Higher orders and four dimensions}

Computing higher orders in the Laurent expansion is interesting for
several reasons. First of all, we obtain a result for the four-dimensional
case in this way. While the pole terms of eq. \ref{eq:S(4)} were
already known, we obtain \cite{AdaBogWei3} the coefficient $S^{(0)}(4,t)$
in terms of ${S^{(0)}(2,t)},$ $S^{(1)}(2,t),$
$\frac{\partial}{\partial m_{i}^{2}}{S^{(0)}(2,t)},$
$\frac{\partial}{\partial m_{i}^{2}}S^{(1)}(2,t),$ $i=1,\,2,\,3$
by use of dimension shift relations \cite{Tar1,Tar2}. As $S^{(0)}(2,t)$
is given by eq. \ref{eq:result ellipt dilog}, the missing ingredient
here is $S^{(1)}(2,t).$ This coefficient satisfies a fourth order
differential equation 
\[
L_{1,a}L_{1,b}L_{2}S^{(1)}(2,t)=I_{1}(t)
\]
where the inhomogeneous part $I_{1}(t)$ involves a polynomial of
the squared masses and $t$, logarithms of the squared masses and
the coefficient $S^{(0)}(2,t).$ The differential operator factorizes
into two operators of first order $L_{1,a},\, L_{1,b}$ and an operator
$L_{2}$ of second order which we already know from eq. \ref{eq:DiffEq y}.
Due to this factorization, we obtain the second order differential
equation 
\[
L_{2}S^{(1)}(2,t)=I_{2}(t)
\]
where the only difference to eq. \ref{eq:DiffEq y} is a more
complicated inhomogeneous part $I_{2}(t).$ Applying variation of
constants as above, we obtain an explicit result for $S^{(1)}(2,t)$
which we express in terms of E-functions. We arrive at \cite{AdaBogWei3}
\begin{eqnarray*}
S^{(1)}(2,t) & = & \frac{\psi_{1}(q)}{\pi}E^{(1)},\\
E^{(1)} & = & \left(\sum_{j=1}^{3}\left(\textrm{E}_{1;0}(w_{j};1;-q)-\frac{1}{3}E_{1;0}(w_{j};-1;-q)\right)-\frac{2}{3}\sum_{j=1}^{3}\ln\left(\frac{m_{j}^{2}}{\mu^{2}}\right)\right.\\
 &  & \left.-6\textrm{E}_{1;0}(-1;1;-q)\right)E^{(0)}+E_{R}^{(1)}
\end{eqnarray*}
where $E_{R}^{(1)}$ is a linear combination of the functions $\textrm{Li}_{2},$
$\textrm{Li}_{3},$ $\textrm{Li}_{2,1},$ $\textrm{E}_{3;1}$ and
$\textrm{E}_{0,1;-2,0;4}$ and where $\textrm{E}^{(0)}$ is defined in eq. \ref{eq:def E^(0)}. With the help of this result, one obtains
the coefficient $S^{(0)}(4,\, t)$ of the sunrise integral around
four dimensions.

The other reason for computing even higher orders in the Laurent series
is our interest in the functions appearing there. Now we consider
the case of equal masses $m=m_{1}=m_{2}=m_{3}$. As a first step,
we define 
\[
\tilde{S}(2-2\epsilon,t)=\sum_{j=0}^{\infty}\epsilon^{j}\tilde{S}^{(j)}(2,t)
\]
by
\[
S(2-2\epsilon,t)=\Gamma\left(1+\epsilon\right)^{2}\left(\frac{3\mu^{4}\sqrt{t}}{m(t-m^{2})(t-9m^{2})}\right)^{\epsilon}\tilde{S}(2-2\epsilon,t).
\]
The differential equations simplify for $\tilde{S}$ and allow us
to recursively express any coefficient in the Laurent series as
\[
\tilde{S}^{(j)}=-\frac{\psi_{1}}{\pi}\int_{q_{0}}^{q}\frac{dq_{1}}{q_{1}}\int_{q_{0}}^{q_{1}}\frac{dq_{2}}{q_{2}}\left(a_{j}+b\tilde{S}^{(j-2)}\right)\textrm{ for }j\geq2.
\]
We can show that all functions $a_{j}$ and the $j$-independent function
$b$ in this equation can be expressed as products of ELi-functions.
The lowest coefficients $\tilde{S}^{(0)},$ $\tilde{S}^{(1)}$ are
immediately obtained from our previous results in terms of E- or ELi-functions.
Therefore, due to the properties of our classes of functions discussed
in section \ref{sub:Definitions}, all coefficients $\tilde{S}^{(j)}$
can be expressed in terms of E-functions, together with classical
and multiple polylogarithms. As a consequence, the same is true for
all orders of $S(2-2\epsilon,t)$ \cite{AdaBogWei5}.

\section{Conclusions\label{sec:Conclusions}}

By use of the class of functions defined in section \ref{sub:Definitions}
we have computed the sunrise integral to order $\mathcal{O}(\epsilon)$
in two dimensions and to order $\mathcal{O}(\epsilon^{0})$ in four
dimensions. For the case of equal masses and two dimensions, we presented
an algorithm to compute all orders. The iterated integral structure
of our class of functions has shown to be useful in the systematic
use of the method of differential equations. It therefore suggests
itself to be used in future computations of further integrals, beyond
multiple polylogarithms and beyond the sunrise.

\end{document}